\documentclass{optica-article}

\journal{opticajournal} 

\articletype{Research Article}
\graphicspath{ {./Figures/} }
\usepackage{lineno}
\usepackage{graphicx}
\usepackage{dcolumn}
\usepackage{bm}
\usepackage{caption}
\usepackage{subcaption}
\usepackage{gensymb}
\usepackage{xcolor}
\usepackage{wrapfig}

\begin{document}

\title{Holographic imaging of antiferromagnetic domains with in-situ magnetic field}

\author{Jack Harrison \authormark{1}, Hariom Jani \authormark{1,2,*}, Junxiong Hu \authormark{2}, Manohar Lal \authormark{3}, Jheng-Cyuan Lin \authormark{1}, Horia Popescu \authormark{4}, Jason Brown \authormark{1}, Nicolas Jaouen \authormark{4}, A. Ariando \authormark{2}, Paolo G. Radaelli \authormark{1,*}}

\address{\authormark{1} Clarendon Laboratory, Department of Physics, University of Oxford, Oxford, OX1 3PU, UK\\
\authormark{2} Department of Physics, National University of Singapore, Singapore\\
\authormark{3} Department of Electrical and Computer Engineering, National University of Singapore, Singapore \\
\authormark{4} Synchrotron SOLEIL, L’Orme des Merisiers, Saint-Aubin, B.P. 48, 91192 Gif-sur-Yvette, France}

\email{\authormark{*} hariom.jani@physics.ox.ac.uk, paolo.radaelli@physics.ox.ac.uk} 


\begin{abstract*} 
Lensless coherent x-ray imaging techniques have great potential for high-resolution imaging of magnetic systems with a variety of in-situ perturbations. Despite many investigations of ferromagnets, extending these techniques to the study of other magnetic materials, primarily antiferromagnets, is lacking. Here, we demonstrate the first (to our knowledge) study of an antiferromagnet using holographic imaging through the ‘holography with extended reference by autocorrelation linear differential operation’ technique. Energy-dependent contrast with both linearly and circularly polarised x-rays are demonstrated. Antiferromagnetic domains and topological textures are studied in the presence of applied magnetic fields, demonstrating quasi-cyclic domain reconfiguration up to 500 mT.
\end{abstract*}

\section{Introduction}

Synchrotron-based x-ray imaging has become an invaluable tool for the study of magnetic, quantum and functional materials for a wide variety of fundamental and application-based research. X-ray photoelectron emission microscopy (X-PEEM), scanning transmission x-ray microscopy (STXM) and several other related techniques have been successfully employed to study a wide range of materials with spatial resolutions down to a few tens of nm \cite{XRayImagReview, DWLogic, Francis, BFO}. All of these methods rely on core-level spectroscopy to produce contrast, but differ in the way the image is created. The availability of appropriate sample environments to apply \textit{in-situ} perturbations to the sample is of great importance to enhance the impact of this research, and much progress has been made to provide non-standard environments and stimuli \cite{GasCell1, CuMnAs, ThIG_Bubbles}.

In this respect, many x-ray based imaging techniques suffer from fundamental limitations, which restrict the complexity of the sample environment and consequently the accessible phase space. For transmission-based techniques such as STXM, the limitation is geometrical and relates to the requirement for a zone plate very close to the sample \cite{XRayImagReview}. In the case of X-PEEM, large voltage differentials at the sample stage are required to extract and accelerate secondary electrons, significantly limiting the available space around the sample \cite{PEEM}. Moreover, the very nature of the X-PEEM technique, which is based on charged particles, makes it difficult to image samples in large applied magnetic or electric fields. As an alternative, photon-based lensless imaging is particularly appealing, since the x-ray beam is formed far away from the sample stage and no distortion is introduced when fields are applied \cite{COMET}. The necessity to employ coherent x-ray beams, traditionally regarded as a limitation of these methods, is being progressively overcome by modern synchrotron designs, which provide large coherent fractions in the soft x-ray regime. A further boost to techniques based on bright coherent sources will be provided by fourth-generation storage rings based on the multi-bend achromat lattice concept \cite{SynchReview, Synchro4}, as well as x-ray free electron lasers \cite{HoloFEL}. In this context, it is worth emphasising that full-field lensless methods do not require scanning, and are therefore ideal for time-resolved studies.

X-ray Fourier transform holography (FTH) is a family of related lensless imaging techniques, all aiming to reconstruct a real space, full-field image of a sample by performing an inverse Fourier transform on a measured interference pattern between the sample and a reference structure \cite{HoloReview}. Unlike STXM and related techniques such as ptychography \cite{PtychoReview, XRayImagReview}, FTH does not rely on close-in imaging optics. This creates greater flexibility for sample environment and external stimuli \cite{COMET}, and also means that the fundamental resolution of the technique is limited by the x-ray wavelength and precision of the reference structure, rather than the resolution of the employed optics, and can therefore be better by up to an order of magnitude. The main drawback of FTH is the need to create an x-ray mask, including a reference with feature sizes that are comparable to the resolution one wants to obtain, which requires a significant investment in nanofabrication. 

Within the FTH family, holography with extended reference by autocorrelation linear differential operation (HERALDO) \cite{Holo1, Holo2} is a variant of standard holographic techniques that uses an extended shape (such as a slit) as the scattering reference that interferes with the diffraction from the object under study. As above, both the sample and the reference are defined by a strongly x-ray absorbing mask, into which an object hole and the reference object are cut. A suitable linear differential filter is applied to a measured far-field interference pattern, such that the inverse Fourier transform of the resulting diffraction pattern generates several real-space reconstructions of the object (see Fig. \ref{fig:HoloMask}a) \cite{XHolo, HoloBiskyrm, HoloMnNiGa, HoloTime}. HERALDO has several advantages over standard FTH. Firstly, the flexibility to manufacture general shaped references rather than the small holes required by standard FTH allows for enhanced flexibility in sample geometry and manufacturing processes \cite{Holo1, Holo2}. As the extended references have greater x-ray transmission than hole references for FTH, lower photon fluxes are required for comparable image quality. Finally, the resolution of HERALDO is determined by the size of the sharpest corner one can manufacture, rather than the overall size of the reference hole, meaning that it can be in principle up to an order of magnitude finer \cite{XHolo}. Since its original development, HERALDO has been mainly employed to image ferromagnetic materials through circular dichroism \cite{XHolo, HoloBiskyrm, HoloMnNiGa, HoloReview, HoloTime}. As discussed further below, HERALDO images can be created by exploiting many other contrast mechanisms, and it is therefore suitable to study a much wider class of materials than conventional ferromagnets.

\subsection{Magnetic and topological textures in $\alpha$-Fe$_2$O$_3$}

$\alpha$-Fe$_2$O$_3$ is a quasi-collinear canted antiferromagnet (AFM) which has been studied for decades as a model magnetic system. More recently, it has been identified as a candidate for applications in spintronics, due to its long-range spin diffusion, reasonably large spin-Hall magnetoresistance, low Gilbert damping and widely tunable magnetic anisotropy \cite{SpDiff, SpTrans, SpHall1, SpHall2, HDoping, MRS}. $\alpha$-Fe$_2$O$_3$ has a high N\'eel temperature ($T_\text{N}\approx$ 950\,K in bulk samples); its topological properties stem from the unusually low magnetic anisotropy, which changes sign from in-plane (IP) to out-of-plane (OOP) upon cooling through the so-called the Morin transition ($T_\text{M}\approx$ 260\,K in bulk samples) \cite{MorrishBook, Morin}. Other than very close to $T_\text{M}$, the basal-plane anisotropy is orders of magnitudes smaller than the axial anisotropy, meaning that the different in-plane spin directions are almost identical energetically. This favours complex in-plane spin patterns, which are strongly reminiscent of Schlieren textures in liquid crystals and can have similar topological characters. Moreover, $\alpha$-Fe$_2$O$_3$ has a weak ferromagnetic moment above the Morin transition due to a small in-plane canting of the two sublattices towards each other, thereby allowing the antiferromagnetic domains to weakly couple to an applied field \cite{MorrishBook, Morin}.

Very recently, interest in $\alpha$-Fe$_2$O$_3$ has been rekindled thanks to the discovery that this system can host topological magnetic textures (merons, antimerons and bimerons) at room temperature, making it a potential platform for racetrack-based spintronic devices and other unconventional computing paradigms \cite{Hariom, Francis, MySims, MRS}. These studies exploited energy-dependent linear-dichroic x-ray absorption (referred to as energy-XMLD contrast) to image antiferromagnetic domains and topological textures in both thin films \cite{Francis, Hariom} and free-standing membranes \cite{STXM}. We demonstrated therein that topological textures can be repeatedly nucleated by a first-order analogue of a Kibble-Zurek transition \cite{Kibble1, Zurek, KZMech}. In spite of the promising results obtained on $\alpha$-Fe$_2$O$_3$ by X-PEEM and STXM, a bottleneck in research has been the inability to apply steady-state in-situ perturbations, such as magnetic fields, in combination with imaging. For example, although we were able to demonstrate that sufficiently large \textit{ex-situ} magnetic fields result in bimeron annihilation whilst preserving the non-topological characters of the AFM domain structure \cite{Hariom}, we were not able to image the sample whilst this process was unfolding due to the aforementioned limitations of the technique. Such \textit{in-situ} field investigations are crucial as our recent study has unveiled that topological AFM textures in $\alpha$-Fe$_2$O$_3$ host monopolar, dipolar and quadrupolar \textit{emergent} magnetic charge distributions coupled to the AFM order \cite{NV}, enabling their direct manipulation via Zeeman interaction.

Following our recent success in growing free-standing $\alpha$-Fe$_2$O$_3$ membranes hosting topological textures, which are suitable for imaging in transmission \cite{STXM}, we are now able to overcome this limitation using the HERALDO scheme in energy-XMLD contrast mode. In this paper, we employ HERALDO to image antiferromagnetic domain structures (for the first time, to our knowledge), including topological textures, in $\alpha$-Fe$_2$O$_3$. We employ both energy and linear dichroic (XMLD) contrast to image the full director field associated with the spatial variation of the N\'eel vector. We also demonstrate that these textures can be manipulated by \textit{in-situ} application of steady-state magnetic fields up to 800\,mT, both above and below $T_\text{M}$, further emphasising the role that FTH methods can play in the study of extended phase diagrams. Furthermore, we show that circularly polarised x-rays can be used to produce HERALDO images of magnetic antiphase domain walls (ADWs) below $T_\text{M}$, something that is generally considered impossible for antiferromagnets.

\section{Materials and Methods}

\subsection{ \label{sec:Sample} Sample Preparation}

\begin{figure*}
\centering
\includegraphics[width=\textwidth]{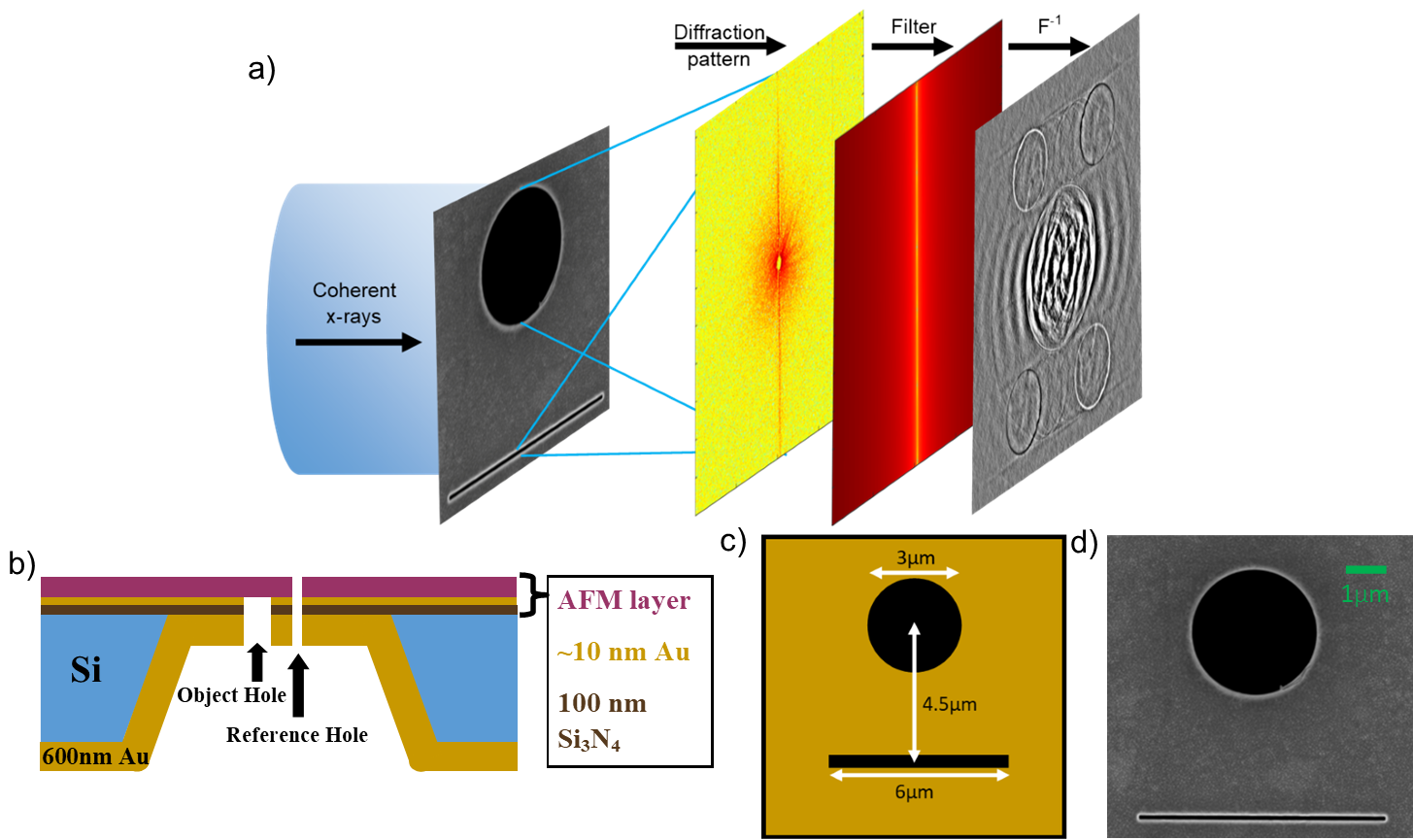}
\caption{a) Schematic of HERALDO technique, wherein a coherent X-ray beam illuminates an object hole and slit, leading to an interference pattern. The application of a linear filter and recreation of the final image through an inverse Fourier transform are also shown. b) Cross-section diagram of our masks (not to scale) with sample post-transfer, highlighting the various layers and the milling structure. c) Diagram of object hole and reference slit layout with relevant dimensions. d) Scanning electron microscope image of an example hole/slit in our holography mask after preparation in FIB but before sample application.}
\label{fig:HoloMask}
\end{figure*}

The resolution of HERALDO is primarily determined by the size of the reference corners that correspond to the reconstructions, i.e., how carefully and precisely one can prepare the mask. In addition, as the mask will never be perfectly x-ray absorbing, the overall contrast of the technique is determined by the relative x-ray transmission of the object relative to the mask. There are also a set of criteria that determine the relative object-reference sizes and spacing to ensure that different reconstructions do not overlap \cite{Holo1, Holo2}. This effectively shifts the experimental difficulties usually associated with beamline operation to the sample and mask preparation, thereby allowing for greater flexibility in the experimental chamber itself \cite{COMET}.

To prepare the mask, we used commercial (Silson) holders made of 100\,nm thick Si$_3$N$_4$ membranes mounted on 5\,mm x 5\,mm x 0.5\,mm Si substrates with a 1\,mm x 1\,mm window cut out as a base. The mask was grown via e-beam evaporation of Au wire using an Edwards EB3 electron gun. The chamber was evacuated to a base pressure of $1.2 \times 10^{-5}$\,mbar. A 600\,nm Au layer was evaporated at a rate of 0.25\,nm/s using a 55\,mA beam current. As Au has a typical x-ray attenuation length of 60\,nm in the energy range of interest (700-800\,eV) \cite{AttenCalc}, we expect a mask transmission of approximately 1/2000 of the incident intensity, which should be sufficient to ensure a reasonable contrast. Moreover, about $\approx$10\,nm of Au was coated on the reverse of the membrane to make the surface conductive for the further processing steps described below.

After Au coating, the masks were mounted in a FEI/ThermoFisher Nova 600 NanoLab DualBeam FIB/SEM to perform the milling. An ion beam voltage of 30\,kV and beam current 28\,pA was used to mill both the slits and holes. These values were optimised to make the slits as narrow and uniform as possible in order to minimise spurious reconstructions and maximise attainable resolution. We milled a 10x3 array of 3\,$\mathrm{\mu m}$ diameter holes separated by 50\,$\mathrm{\mu m}$. Along with each hole, an associated slit was milled 4.5\,$\mathrm{\mu m}$ from the centre of the hole. Slits were 6\,$\mathrm{\mu m}$ long and slightly off-centre (to aid with the filter application algorithm). These are shown schematically and as imaged in a scanning electron microscope in Fig. \ref{fig:HoloMask}b-d. These dimensions ensure that all the reconstruction separation criteria are satisfied. The milled slits had a width between 40\,nm and 70\,nm, indicating that this is the best resolution we would expect to achieve. This is confirmed in Fig. \ref{fig:LT}a,b, wherein linecuts of example domain walls suggest a resolution close to 50\,nm as will be explained further in section \ref{sec:Resolution}. The array of multiple holes allowed us to explore different regions of the sample, helping to overcome the limited field of view that is otherwise a downside of this technique.

(001)-oriented, 30\,nm thick $\alpha$-Fe$_2$O$_3$ membranes were prepared by pulsed laser deposition growth on SrTiO$_3$ substrates with a water-soluble Sr$_3$Al$_2$O$_6$ (SAO) layer and a (111)-oriented SrTiO$_3|$LaAlO$_3$ buffer to enhance the crystalline quality of the films. These membranes were transferred onto the prepared sample holders by using a lift-off and targeted transfer technique after water-etching the SAO layer; for full details of the preparation process and sample characterisation see \cite{STXM}. The targeted transfer technique used herein minimised bent or crumpled regions and the object holes used for imaging were confirmed to be absent of large structural defects via optical imaging (see Fig. S1). The buffer layer remains attached to the $\alpha$-Fe$_2$O$_3$ post-transfer, but has no effect on the magnetic properties of the film \cite{STXM}. To re-mill the slits, we used a focused beam of Neon ions in a Zeiss Orion NanoFab (He/Ne ion Microscope). We used a beam current of 5\,pA, ion beam energy of 26\,KeV, a dose of 3.5x10$^{17}$\,ions/cm$^2$, and a fixed dwell time of 25\,µs. An area 7\,$\mathrm{\mu m}$ x 0.4\,$\mathrm{\mu m}$ was milled about each slit in a raster scan with a beam spacing 1\,nm x 1\,nm. This removed all of the sample and buffer layer covering the slits, thereby ensuring that x-rays are directly transmitted through the reference as necessary for a clean reconstruction.

\subsection{HERALDO contrast mechanism with XMLD and energy difference}

As with all FTH techniques, HERALDO reconstructs the full complex transmittance of a sample from a measured interference pattern between an object and some reference structure \cite{HoloReview}. This is done by applying a suitable filter to the diffraction pattern and then taking the inverse Fourier transform \cite{Holo1, Holo2, XHolo}, as shown schematically in Fig. \ref{fig:HoloMask}a.

In general, the energy-dependent optical index of a sample can be written as \cite{HoloPhase}
\begin{equation}
n(\omega) = 1 - \delta(\omega) + i\beta(\omega).
\end{equation}
The real ($\delta(\omega)$) and imaginary ($\beta(\omega)$) parts of the optical index can be related to the standard scattering factors $f'(\omega)$ and $f''(\omega)$ via the relations
\begin{equation}
\delta(\omega) = \frac{r_0 \lambda^2}{2\pi}\rho(Z+f'(\omega)) \text{ and } \beta(\omega) = \frac{r_0 \lambda^2}{2\pi}\rho f''(\omega).
\end{equation}
Here, $r_0$ is the classical electron radius, $\lambda$ is the wavelength of the incident x-rays, $\rho$ is the atomic number density, and $Z$ is the number of atomic electrons.

The complex transmittance of a thin sample of thickness $d$ at wavelength $\lambda$ and with polarisation $\alpha$ is therefore
\begin{equation}
T_\alpha(\omega) = \mathrm{exp}\left[\frac{2\pi i}{\lambda} \int_0^d (n_\alpha(\omega)-1)\,dz \right] = \mathrm{exp}\left[-\frac{2\pi d\beta_\alpha(\omega)}{\lambda} \right]\mathrm{exp}\left[-\frac{2\pi di\delta_\alpha(\omega)}{\lambda} \right],
\end{equation}
where we have assumed that the optical index doesn't vary through the sample thickness --- this is a good approximation in our case, given the thickness of our films. Near an absorption resonance peak both the absorbative (real) and dispersive (imaginary) parts of the transmittance increase, but tend to peak at slightly different energies ($\hbar \omega$) \cite{HoloPhase, HoloLin1}. Furthermore, the different components of the transmittance acquire a dependence on the x-ray polarisation determined by the relative orientation of the x-ray polarisation and the magnetic moment; this defines the magnetic contrast mechanism. For antiferromagnets, such as $\alpha$-Fe$_2$O$_3$, there is a well-known relationship between the orientation of the local N\'eel vector and the x-ray absorption, namely $\propto \cos(2\theta)$ with $\theta$ the angle between the N\'eel vector and x-ray polarisation axis \cite{FeXAS, Fe3O4_XMLD, Hariom}. As the real and imaginary parts of the optical index are related by the Kramers-Kronig relation, the dispersive part of the transmittance will have the same polarisation dependence. Hence, HERALDO images reconstructed from either the amplitude or the phase of the transmittance will directly map the local N\'eel vector orientation up to a sign, as with previous studies based only on absorbative imaging \cite{Francis, Hariom, STXM}. The real and imaginary parts of the transmittance will also contain similar information, though with a different $2\theta$ dependence. It is important to note that the difference in peak energies for absorbative vs dispersive contrast means that `real' and 'imaginary' images cannot be simultaneously optimised, so that only one of them will have optimum contrast. The amplitude or phase of the reconstruction may also be sensitive to particular features, depending on the case.

Energy-XMLD (E-XMLD) images are constructed with fixed linear x-ray polarisation in the plane of the sample with normal incidence, but with diffraction patterns taken at two different energies near the $L_3$ Fe peaks. This works because the dichroic signal reverses sign several times across the spectrum, and one can choose energies that maximise this difference \cite{FeXAS, Fe3O4_XMLD, Hariom}. The contrast is defined in terms of the intensity at two energies, $I_\text{E$_1$}$ and $I_\text{E$_2$}$ as
\begin{equation}
I_\text{E-XMLD} = I_\text{E$_1$} - I_\text{E$_2$},
\end{equation}
referred to generally as a difference image. As the HERALDO processing is linear, one can perform the subtraction either before or after applying the filter and performing the inverse Fourier transform. Furthermore, the background intensity of each diffraction pattern is separately normalised prior to subtraction. This process ensures that images taken with HERALDO are analogous to those we studied previously through X-PEEM \cite{Francis, Hariom} or STXM \cite{STXM}, but with the added caveat that the primary contrast may come from either the real or imaginary part of the reconstruction. In our case, the energies chosen here were optimised to maximise the contrast in the \emph{real} part of the HERALDO reconstruction.

Beyond the E-XMLD images discussed above, by rotating the x-ray polarisation within the sample plane, one can exploit the fact that the local contrast $I_\text{E-XMLD} \propto \cos(2\theta)$ ($\theta$ being the angle between the local magnetisation and the x-ray polarisation direction) to reconstruct a full 2D map of the local N\'eel vector orientation up to a sign (N\'eel director field), referred to throughout as a vector map. Due to the $\cos(2\theta)$ periodicity, a set of E-XMLD images collected between 0\textdegree~ and 90\textdegree~ is in principle sufficient to achieve the full reconstruction. Our implementation is similar to that performed previously using absorption-based techniques \cite{Francis, Hariom, STXM}, but with the additional caveats discussed above. 

\subsection{Instrumentation and measurements}

X-ray holography experiments were performed on the COMET instrument at the SEXTANTS beamline, SOLEIL. Far field diffraction patterns formed via the interference of the sample and reference aperture were recorded using either a 2048x2048 pixel 16-bit PI-MTE CCD camera (Princeton instruments) or a new 2048x2048 pixel sCMOS (AXIS Photonique) \cite{HoloCamera}. The X-ray polarization was controlled with an HU44 APPLE II undulator. Magnetic fields were applied to the sample using a tunable array of 4 permanent magnet (NdFeB) rods \cite{COMET}. The sample temperature was controlled with a commercial He cryostat/heater (Janis, USA) connected to the sample stage with a local flexible stripe, in principle allowing for sample temperatures 20\,K-800\,K \cite{COMET}, although only temperatures 280\,K-320\,K were used here. A Lakeshore 350 controller performs the temperature regulation with stability better than 100\,mK. Post-processing of measured diffraction patterns, such as the application of the linear filter and inverse Fourier transform, were performed using a custom matlab script. Vector maps were created from a set of ten images with different polarisation angles using a custom python script. For low-temperature vector maps ($T<T_\text{M}$), OOP regions were masked out by applying a binary mask to the angular-averaged E-XMLD image, with a consistent threshold across images at the same location.

\section{Results and Discussion}
\subsection{\label{sec:LT} Imaging domain walls below $T_\text{M}$}

\begin{figure}
\centering
\includegraphics[width=\textwidth]{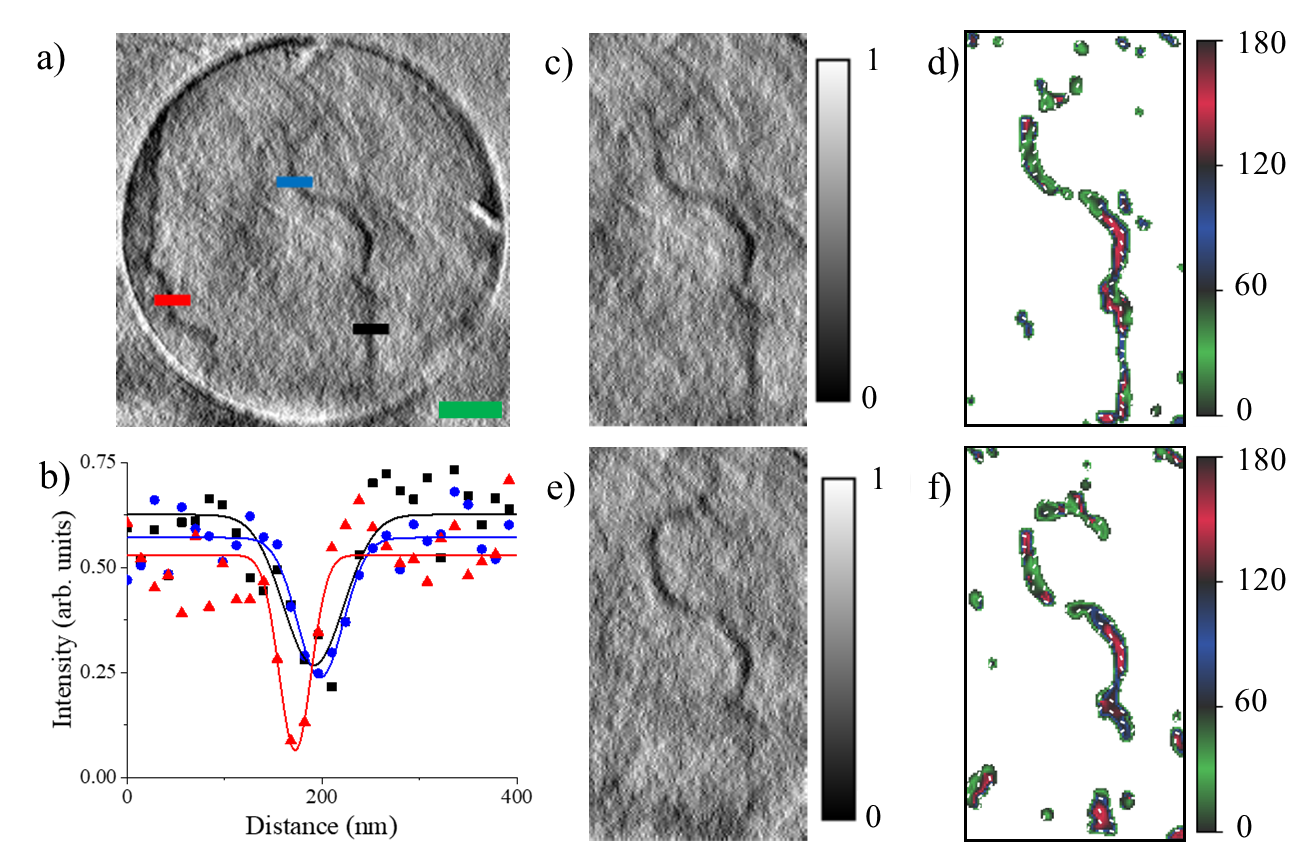}
\caption{a) Holographic image of an AFM domain wall and (b) a set of linecuts taken across that domain wall along with corresponding Gaussian fits of the intensity, demonstrating the resolution achieved with this technique. c,e) Holographic E-XMLD images and (d,f) vector maps taken at 280\,K (below $T_\text{M}$) across the ADW shown in (a). (c,d) are in the absence of an applied field and (e,f) are with a constant 650\,mT field applied towards the right of the image. Green scale bar in (a) is 500\,nm. Red-green-blue colours and white bars represent fitted N\'{e}el vector orientation, with the angle measured relative to the vertical axis shown by the scale bar.}
\label{fig:LT}
\end{figure}

For $T<T_\text{M}$, $\alpha$-Fe$_2$O$_3$ hosts a set of AFM domains oriented along the $c$-axis, related by time-reversal symmetry and connected by in-plane antiphase domain walls (ADWs) \cite{Hariom}. For our membranes, the $c$-axis is oriented OOP of the films, and as such the domains at this temperature are oriented along the x-ray k-vector in normal-incidence geometry. Therefore, E-XMLD contrast allows us to clearly distinguish between the OOP domains and the ADWs, although the pair of time-reversed OOP domains have the same contrast and so are indistinguishable. 

\subsubsection{\label{sec:Resolution} Determination of the resolution of the mask}

Since the domain walls are narrow and their width is known from previous studies, analysis of the ADW images obtained with $T<T_\text{M}$ enables us to assign an upper limit to the `instrumental resolution' obtained with the masks and samples as prepared here, as well as with the specific experimental configuration employed. To this end, we took a set of linecuts perpendicular to domain walls imaged below $T_\text{M}$ and performed a Gaussian fit of the pixel-by-pixel intensity, as shown in Fig. \ref{fig:LT}a,b. Each pixel corresponds to $\sim$14\,nm in real space after the reconstruction, as the 3\,$\mathrm{\mu m}$ diameter holes are found to be 210 pixels wide. The extracted width of the Gaussian fit to each line cut was between 40\,nm and 60\,nm in each case, with larger widths found across some sections of the domain wall. This is consistent with typical domain wall sizes seen below the transition during previous X-PEEM experiments and through micromagnetic simulations \cite{Hariom, MySims}. Thus, the mask preparation resulted in a resolution of approximately 50\,nm, which corresponds closely to the width of the reference slits as extracted from SEM images, similar to Fig. \ref{fig:HoloMask}d. The 14\,nm pixel size results from the lateral dimensions of the camera used and the separation between the sample and detector, as the scattering vector corresponding to the furthest pixel from the image centre as well as the pixel density will also ultimately limit the instrument resolution. These parameters can in principle be optimised over and above what was used in these experiments and are usually less relevant for determining the ultimate resolution of the technique than the size of the reference feature, given our current manufacturing capabilities.

\subsubsection{Field-dependent behaviour of the domain walls below $T_\text{M}$}

Fig. \ref{fig:LT}c shows an example ADW separating two OOP domains, with the associated vector map in Fig. \ref{fig:LT}d. As discussed in our previous work, the domain walls are not homochiral and have a spatially-varying in-plane orientation \cite{Hariom}. The ADWs also vary slightly in width, likely due to local fluctuations in the anisotropy from inhomogeneous strain caused by defects or imprinted serendipitously during membrane transfer \cite{STXM}; nonetheless, we find that the average width is consistent with our previous observations in attached thin films \cite{Hariom}. In addition, some small IP islands are also visible in the vector map, as often seen in such samples when the system is below $T_\text{M}$, resulting from the relatively weak anisotropy of the system close to the Morin transition.

Under application of magnetic fields up to 650\,mT along the IP direction, the ADWs are seen to be structurally robust to change (Fig. \ref{fig:LT}e). This could be partly due to the higher density of defects present in $\alpha$-Fe$_2$O$_3$ membranes relative to their thin-film counterparts \cite{STXM}. Moreover, as the fraction of the ADW that is fully in plane is relatively small, the canted moment and associated coupling with the external field is expected to be fairly weak. Nonetheless, by analysing the vector map of the ADW (Fig. \ref{fig:LT}f) at 650\,mT compared to the zero field case, we observe some reorientation of the IP component of the ADW. This suggests that the coupling is non-zero, and that higher fields could potentially cause complete reorientation within the ADW. Whilst this would not enforce homochirality, as the orientation relative to the domain wall vector would still spatially vary as the wall itself wound through the sample, spin reorientation within ADWs could have potential uses for racetrack-type devices, wherein ADWs perpendicular to the long axis of the racetrack are favoured. We emphasise here that observing such AFM ADW reorientation in the presence of a steady-state in-situ magnetic field is significantly easier in holography compared to other synchrotron-based x-ray techniques, due to the absence of focusing objectives and larger space in the experimental chamber.

\subsubsection{Imaging antiphase domain walls with circularly polarised x-rays}

\begin{figure}
\centering
\includegraphics[width=\textwidth]{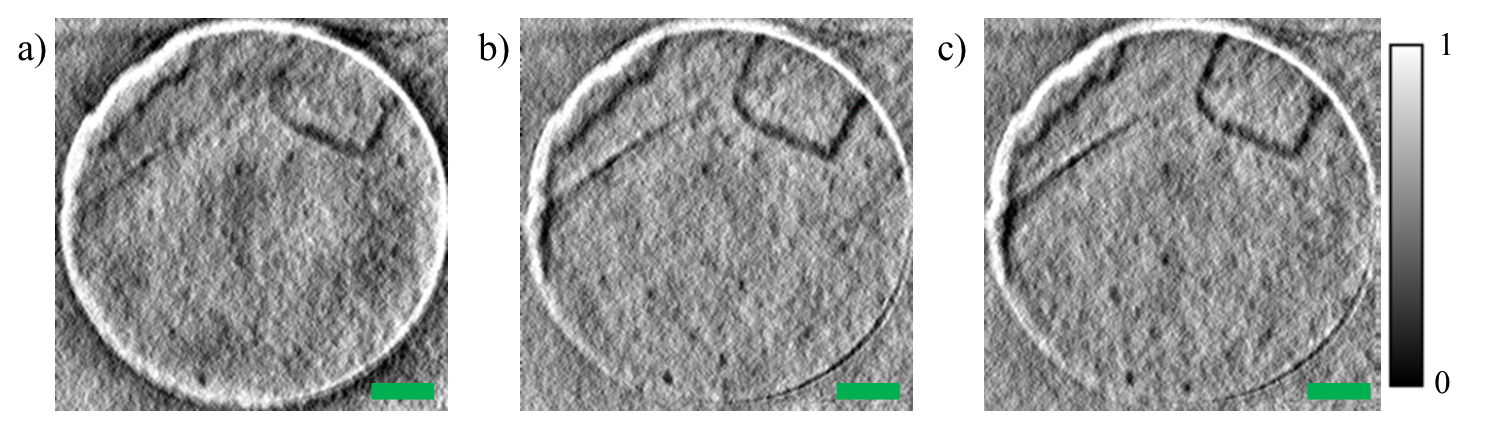}
\caption{a)  Holographic E-XMLD image of ADWs for $T<T_\text{M}$. b,c) CXMEC image of the same ADWs using b) circular left and c) circular right polarised x-rays. Green scale bars are 500\,nm.}
\label{fig:XMCD}
\end{figure}

More interestingly, and contrary to what could naively be expected for AFMs, domain walls can be clearly seen for $T < T_\text{M}$ in HERALDO energy-contrast images with \emph{circularly}-polarised x-rays (Fig. \ref{fig:XMCD}). We refer to these as circular x-ray magnetic energy contrast (CXMEC) images. These were taken with both left- and right-handed circular light and at the same pair of energies as the E-XMLD images and show clear contrast between the ADWs and the OOP background. The contrast is strong and very uniform within the ADWs; crucially, it is free from the variations in IP contrast observed with E-XMLD imaging, which we used to construct vector maps. Our observation of strong contrast with circular light may appear counter-intuitive, since it is generally believed that AFM textures can only be observed with linear x-ray polarisation. In fact, there is no violation of the usual rules here, since our CXMEC images were collected with a single photon helicity and are \emph{not} circularly dichroic. Rather, the contrast simply arises from the different transmittance of circular light for OOP vs IP magnetic moments. By symmetry, this imaging methodology is insensitive to the IP direction within the ADWs, which is both an advantage and a disadvantage, the latter obviously being that one cannot construct vector maps. A significant advantage is that the ADW can be seen in its entirety with a single polarisation setting, without the characteristic null-contrast `breaks' observed with E-XMLD (e.g. in Fig \ref{fig:LT}). Overall, CXMEC appears to be a valuable addition to the imaging `toolkit' for antiferromagnets.

\subsection{\label{sec:HT} Imaging domains above $T_\text{M}$}

\begin{figure}[h]
\centering
\includegraphics[width=\textwidth]{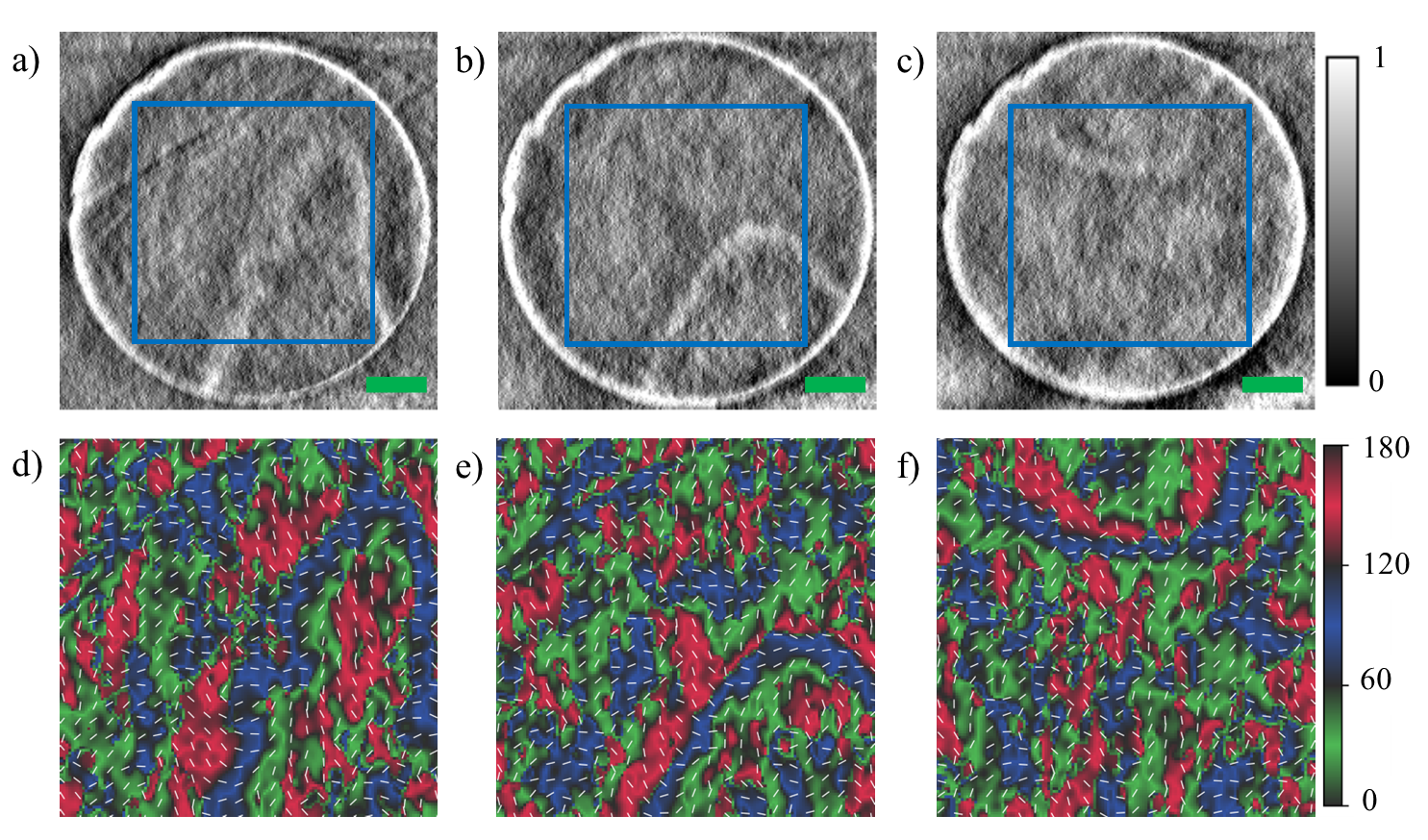}
\caption{(a-c)  Holographic E-XMLD images of AFM domains taken at different field values for the same sample position as Fig. \ref{fig:LT} after warming to 320\,K (above $T_\text{M}$). (d-f) corresponding vector maps taken in the blue boxed regions in (a-c). (a,d) correspond to the zero-field state and (b,e) were taken with a 550\,mT steady-state field applied to the right of the image. (c,f) are taken with the field reversed along the horizontal direction with a strength of 400\,mT. Green scale bars are 500\,nm. The white bars in the vector maps show the local N\'{e}el vector orientation.}
\label{fig:FieldHoloMap}
\end{figure}

After warming the sample through $T_\text{M}$ and up to 320\,K, we expected the ADWs to widen and form an IP matrix of 3-fold domains separated by 120\textdegree ~and their time-reversed counterparts, akin to what was observed previously in thin films \cite{Hariom}. Indeed, our E-XMLD HERALDO images change drastically compared to low temperature. At 320\,K and zero magnetic field, a domain is clearly visible in single-polarisation images (Fig. \ref{fig:FieldHoloMap}a). Performing an E-XMLD vector map by rotating the polarisation of the incident beam allows us to clearly map the three pairs of 120\textdegree ~domains (Fig. \ref{fig:FieldHoloMap}d), and to identify the original domain wall as a narrow `blue' band (N\'eel vector `right' or `left' with respect to the Fig.). As previously explained, each domain is indistinguishable from its time-reversed counterpart, due to the inherent 180\textdegree ~symmetry in this technique. In the centre of the figure, the `blue' domain appears to be pinched and connect two pairs of green/red domains, which are likely to be time-reversed counterparts (green/anti-green and red/anti-red).

 Under application of a 550\,mT IP field along the horizontal direction, the domains changes drastically, as clearly evidenced in the vector map (Fig. \ref{fig:FieldHoloMap}e). In particular, the area enclosed by the narrow `blue' domain is seen to shrink significantly with respect to zero field. Due to beamtime constraints, the magnetic field was applied in steps without any intermediate imaging. Therefore, it is difficult to tell from the vector map alone whether this corresponded to a movement of the `blue' domain towards the bottom right corner or to fracturing into several smaller domains. Moreover, there is a large-scale redistribution of the domains compared to the zero-field case. As the field is applied to the right of the images and the canted moment of a given domain is perpendicular to its N\'eel vector, one would expect `blue' domains to have minimal coupling to the applied field, whereas `green' and `red' domains should both couple equally. Rather than the `blue' domain contracting, therefore, it is more likely that the `red'/`green' domains inside couple negatively with the applied field and so shrink, whereas those outside couple positively and therefore expand. Here, an important observation is that the applied field provides a symmetry-breaker between the time-reversed domain pairs, which have opposite canted moment and therefore are favoured/disfavoured (respectively) in the presence of an applied field. Hence, any newly nucleated domains must have canted magnetic moment magnetic moment parallel to the applied field, whereas the shrunken domains must be their time-reversed counterparts. The blue domains largely serve as pathways for this domain repopulation, as this pair of domains are roughly equal in energy with respect to the applied field. A similar trend (in reverse) is evident when the field direction is reversed (Fig. \ref{fig:FieldHoloMap}c,f). Our data clearly demonstrate that, given enough data collection time, one could reconstruct in detail the evolution of the AFM texture \textit{in situ} as a function of magnetic field, temperature or in indeed other external parameters such as uniaxial/biaxial strain \cite{STXM}.

\subsection{\label{sec:Topo} Topological textures in magnetic fields}

\begin{figure}[h]
\centering
\includegraphics[width=\textwidth]{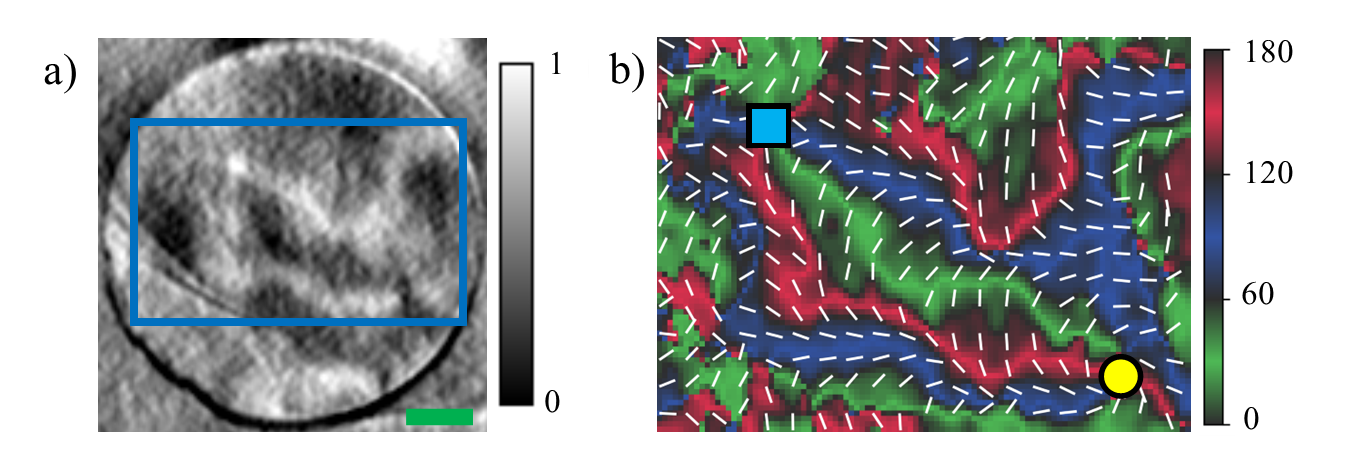}
\caption{a)  Holographic E-XMLD image and (b) corresponding vector map of topological textures in the absence of an applied field. The vector map is taken in the blue boxed region in (a). Green scale bar is 500\,nm.}
\label{fig:HoloMeron}
\end{figure}

Similar to what was observed in our previous studies on attached thin films and freestanding AFM membranes using X-PEEM \cite{Francis, Hariom} and STXM \cite{STXM}, respectively, topological textures were also found to be stable (although seemingly less abundant) in the membranes prepared for this holography experiment. Both a single E-XMLD image and the associated vector map for a pair of topological textures are shown in Fig. \ref{fig:HoloMeron}, taken at 320\,K in the absence of an applied field. A meron and an antimeron, shown by a yellow circle and a blue square respectively, are clearly present in the image, as determined by the winding of red/blue/green domains around the core. The meron and antimeron in the pair are separated spatially by $\approx$ 1.7 $\mathrm{\mu m}$, much further than the magnetic exchange length of the system ($\approx 100$\,nm \cite{MySims}) and therefore they should not be categorised as a bound bimeron system. As for all E-XMLD-based techniques, we are not directly sensitive to the topological charge $\bm{q}$, and we cannot determine whether the pair is trivial ($\bm{q}=0$) or topological ($\bm{q} \pm 1$) \cite{MySims}. Besides providing a direct verification of the results and conclusions of our previous studies \cite{STXM}, the observation of topological textures by HERALDO offers a unique opportunity to study \emph{in situ} the effect of relatively large high magnetic fields on such textures --- something that we were never able to do with SXTM or X-PEEM.

Under application of a 500\,mT magnetic field in the IP direction towards the right (as shown in Fig. \ref{fig:HoloMeron}) to the pair of topological textures, we observe an obvious shrinking of the domains connecting the two cores, as seen in Fig. \ref{fig:Fieldcycling}b compared to the initial zero field state (Fig. \ref{fig:Fieldcycling}a). This appears to be a largely reversible change, since reducing the field down to 100\,mT returns the domains roughly the original position (Fig. \ref{fig:Fieldcycling}c). A further cycle to 500\,mT and back shows an almost identical behaviour (Fig. \ref{fig:Fieldcycling}d,e). This suggests that the main effect that these fields have on topological textures is to alter the energy landscape and thereby modify the proportion of each domain `flavour', whilst preserving the overall winding structure around the topological cores, as discussed below. The almost exact reproducibility under field cycles is probably the result of defect pinning of the topological textures and domain structures, making it more energetically favourable for them to return to their original positions when the applied field is removed. This emphasises the topological protection inherent to the textures, as well as the inherent immunity to applied fields often stated as one of the primary benefits of antiferromagnets for spintronics applications. 

\subsubsection{A qualitative model of a meron-antimeron pair in applied field}

\begin{figure}
\centering
\includegraphics[width=\textwidth]{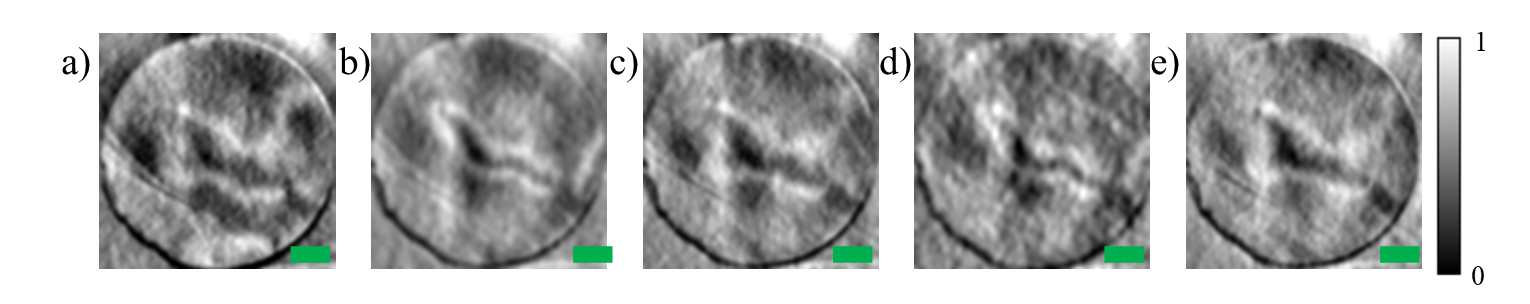}
\caption{Holographic E-XMLD images taken at the position of the pair of topological textures shown in Fig. \ref{fig:HoloMeron} as a function of field, cycling to 500\,mT and back down to 100\,mT twice. The sequence is: (a) 0\,mT initial state, (b) 500\,mT, (c) 100\,mT, (d) 500\,mT, and (e) 100\,mT. The lighter regions correspond to the red domains connecting the two topological cores shown in the vector map above. Green scale bars are 500\,nm.}
\label{fig:Fieldcycling}
\end{figure}

The observed shrinkage of certain domains connecting the meron to the antimeron in applied field is clearly due to the fact that some of the domains are less energetically favourable in the presence of the field and therefore shrink. For example, let us assume that the initial configuration around the meron core is $R-\bar{G}-B-\bar{R}-G-\bar{B}$ as in Fig. \ref{fig:HoloMeron}, with barred domains being the time-reversed counterpart of the unbarred domain ($\bar{R}$= anti-red), and let us also assume that the angular range of each domain is roughly the same in zero field. Let the weak magnetisation of the $R$/$\bar{R}$ domain be close to the positive/negative $x$ axis, respectively (N\'eel vector along the positive/negative $y$ axis). Then, upon application of a magnetic field along the positive $x$ direction, we expect the configuration to become $\bm{R}-\bar{\bm{G}}-B-\bar{r}-g-\bar{B}$, where bold/lowercase symbol indicate greater/lower angular range for a given domain. The effect will be maximal for domains with the weak magnetisation oriented parallel/antiparallel to the field. This qualitative scenario is entirely compatible with our observations in Fig. \ref{fig:Fieldcycling}, and could be further confirmed by performing full vector maps as in Fig. \ref{fig:HoloMeron} at each field.

\section{\label{sec:Summary} Conclusions}
We have presented the first example of holographic imaging of antiferromagnetic domains, utilising energy-contrast x-ray magnetic linear dichroism (E-XMLD) and energy-contrast magnetic imaging with circularly polarised x-rays (CXMEC) in $\alpha$-Fe$_2$O$_3$ membranes. As part of this study, we have produced detailed director-field maps of the N\'eel vector based on E-XMLD as well as sharp antiphase domain wall (ADW) images using CXMEC. Our results confirm the presence of topological textures in $\alpha$-Fe$_2$O$_3$ membranes and demonstrate how these textures evolve in the presence of an applied magnetic field. More specifically, we demonstrated cyclic, field dependent evolution of AFM domains in $\alpha$-Fe$_2$O$_3$ due to the coupling between the field and the canted moment. More broadly, we have shown that E-XMLD/CXMEC HERALDO are promising technique for imaging-based investigations of a wide range of AFM materials with \emph{in-situ} perturbations, including temperature control and magnetic field. This work opens way for investigations of other AFM materials as well as the simultaneous application of additional \emph{in situ} tuning parameters such as electric fields, laser pulses, and uniaxial/biaxial strain.

\begin{backmatter}
\bmsection{Funding}

\bmsection{Acknowledgments}
Work done at the University of Oxford was supported by Engineering and Physical Sciences Research Council (EPSRC) (EP/M020517/1) and the Oxford-ShanghaiTech collaboration project. J.H. was supported by the EPSRC (DTP Grant No. 2285094). H.J. acknowledges the support of Marie Skłodowska-Curie Postdoctoral Fellowship under the UK Research and Innovation Horizon Europe Guarantee Funding (EP/X024938/1). H.J., J.X.H., and A.A. acknowledge the funding support from the Ministry of Education (MOE) Singapore under the Academic Research Fund Tier 2 (Grant No. MOE-T2EP50120-0015), the Agency for Science, Technology and Research (A*STAR) under its Advanced Manufacturing and Engineering (AME) Individual Research Grant (IRG) (Grants No. A1983c0034 \& A2083c0054), and the National Research Foundation (NRF) of Singapore under its NRF-ISF joint program (Grant No. NRF2020-NRF-ISF004-3518). We acknowledge SOLEIL for provision of synchrotron radiation facilities at the SEXTANTS beamline under the projects 20200283 and 20210750.

\bmsection{Disclosures}
The authors declare no conflicts of interest.

\bmsection{Data Availability Statement}
Data that support the findings of this study are available from the corresponding author upon reasonable request.

\bmsection{Supplemental document}

\end{backmatter}

\bibliography{reflist}

\end{document}